\documentclass[twocolumn,twoside,slac]{revtex4}
\usepackage{graphicx}
\usepackage{fancyhdr}
\begin{document}

\title{BaBar Web job submission with Globus authentication and AFS access}

\author{R. J. Barlow, A. Forti, A. McNab, S. Salih}
\affiliation{Dept. of Physics and Astronomy, The University of Manchester,
Manchester, UK}
\author{D. Smith}
\affiliation{Dept of Physics, The University of Birmingham, Birmingham, UK}
\author{T. Adye}
\affiliation{Particle Physics Division, Rutherford Appleton Laborastory, Chilton, Didcot, UK}
\author{ }
\affiliation{On behalf of the BaBar computing group}
\begin{abstract}
We present two versions of a grid job submission system produced for the
BaBar experiment. Both use globus job submission to process
data spread across various sites, producing output which can be combined for
analysis.  The problems encountered with authorisation and authentication,
data location, job submission, and the input and output sandboxes are 
described, as are the solutions.
The total system is still some way short of the aims of enterprises such as
the EDG, but represent a significant step along the way.
\end{abstract}

\maketitle

\thispagestyle{fancy}

\section{INTRODUCTION}

\subsection{The BaBar experiment}

BaBar is a major particle physics experiment\cite{babar-ref} running
at the Stanford Linear Accelerator Center (SLAC), which has already produced 
many billions of events, both real and simulated, in thousands of files. 
These events are processed, re-processed and selected by standard production
jobs. In addition the
collaboration numbers 500+ physicists, many of whom are active in running 
individual analysis jobs.
While certainly smaller than the impending experiments at the LHC, 
BaBar's requirements approach them in several respects, and 
meeting the challenge of BaBar is an invaluable rehearsal for the
challenge of the LHC.

\subsection{BaBar computing}

BaBar data processing 
creates a very high demand for computing resources, both in storage
space and CPU time.  Largely as a consequence of this, the collaboration
has moved from a system which was basically central, with all resources being
provided at the SLAC site, to a distributed model with large and medium
sites (`Tier A' and `Tier C') spread across the USA and Europe.

Grid technology provides the obvious means to make these resources available.
However, unlike the LHC experiments, the BaBar system evolved in the pre-Grid
era, and tools have to be provided which can be used with the current system,
rather than building in Grid concepts from the start.  There is also a  
difference in
timing: we require solutions that can be used today rather than in
2007.

We therefore set out as an exercise to see what could be done with existing Grid
technology, such as Globus and AFS, rather than the more complete but longer
term solutions offered by the EDG \cite{EDG-ref}  and similar projects.  
We encountered the
problems familiar  to other developers
 - authentication, authorisation, data location, the input
sandbox and data retrieval - and were able to solve them, in some cases in
more than one way.

\subsection {The Demonstrator}

The work went through two phases.  The first was a project to provide a
Grid based BaBar system that would act as proof of principle - 
the `Demonstrator'. 
The user runs a Web browser through their desktop or laptop, and for
portability no extra software is required on this platform.
The user selects the data to be analysed according to pre-existing BaBar
criteria, and they are processed by the standard analysis job
(`the WorkBook example')\cite{workbook-ref}.
The {\tt http} server is used for file transport.
Standard ntuples are written, and these are retrieved to the user platform where they
can be analysed using ROOT.

\begin{figure}
\includegraphics[width=65mm]{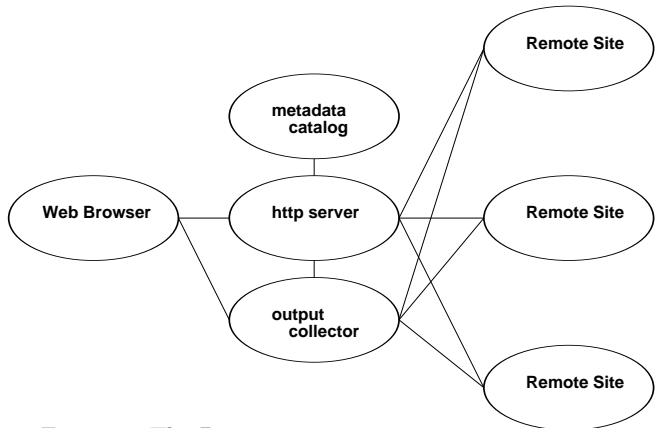}
\caption{The Demonstrator}
\label{f1}
\end{figure}

The Demonstrator was successfully used at eScience events in the summer of 2002
\cite{sheffield-ref}.

\subsection {The {\tt gsub} command}

  The 
second phase 
is a command line system - {\tt gsub}, which gives
easy flexibility for the typical physics analyst.  
The user is presumed to have 
the standard BaBar environment (which enables compilation, data location, etc)
running on their platform.  
AFS is used for the input and output sandboxes, and the tokens are
maintained using the {\tt gsiklog} client and {\tt gsiklogd} server\cite{gsiklog-ref}, so that {\tt gsiklogd} must be
running as a server on the user home system.
This places demands on the user system but not on the remote sites - a
reasonable scenario given that the user is presumed to want to make use of
resources at the remote sites, whereas the remote sites 
have no great incentive to 
provide special facilities for the user.

\begin{figure}
\includegraphics[width=65mm,height=60mm]{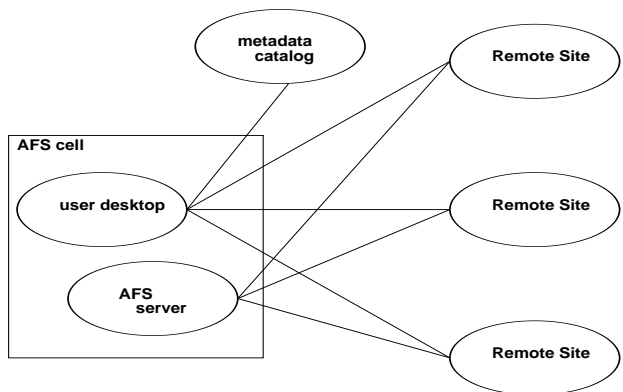}
\caption{The {\tt gsub} command}
\label{f2}
\end{figure}

We describe the two systems, their good features
and also those features which we hope to improve by incorporation of the
more sophisticated middleware currently being written.

\subsection {`BaBarGrid'}

Developments involved a number of clusters of PCs running linux. Several
of these were at UK institutes (Manchester, Birmingham, Rutherford, Imperial,
Bristol, Liverpool, QMUL and RHUL) and comprised identical 40 node PC farms.
Facilities at IN2P3 and Dresden were also included at some stage.
There was good communication and co-operation between the sites and their 
system managers.
 
\section{THE THREE A'S}

\subsection{Authentication}

Standard Grid certificates were used for authentication. Sites involved
had a mutual policy of recognising those authorities accepted by the EDG
and this covered the UK and French institutes. The DOE and German
authorities came into being during this period; before they were available
IN2P3 offered the facility for BaBar users to acquire a certificate.

We resisted early calls for BaBar itself to become a CA.  By maintaining a
clear distinction between authentication and authorisation we realised that
the proper r\^ ole of the experimental 
organisation was with the latter and not the former.

\subsection{Authorisation}

For an authorisation system we set up a VO (Virtual Organisation) \cite{VO-ref}
for BaBar. This is maintained at Manchester and published 
using the {\tt ldap} system in the usual way.

The BaBarGrid sites are available to all members of BaBar. In some cases
(Rutherford and IN2P3, the Tier A sites) this means equal access 
for all members. The Tier C university sites require a system whereby in principle
(or even in practice) priority can be given to local members: institutes
are responsible for the cost and maintenance of their sites, and while they
are generally happy for them to be put to use by outside members, they
do not want their local machines taken over by heavy outside use to an extent
that impinges on the productivity of locals.

A `member of BaBar' has a precise definition \cite{membership-ref}.  
All members have
an account at SLAC, and this has specific `BaBar' authorisation on the 
AFS acl list.  Obtaining this requires user signatures, consent
by the SLAC computer center, authorisation by the group leader, and the
list is kept up to date to weed out members who have moved on elsewhere.
It makes sense to use this existing well-audited system rather than involving 
users in further paperwork.

Any BaBarGrid user has also, by definition, got a Grid Certificate. 
We therefore set up a system whereby all a user has to do to
be entered in the BaBar VO is to copy the DN (Distinguished Name)
from their Grid certificate into a specifically named file in their home area
at SLAC.  A {\tt cron} job scans for such files, checks the (SLAC) userid 
against the AFS acl list, and forwards the DN to the VO server at Manchester.

A second {\tt cron} job runs at each participating site, and picks up the
list of authorised users from the VO.  The local system manager retain control
over this job and can readily modify it, if necessary, to remove any known 
rogues. This has not been necessary and we do not expect it to be, but it
is a useful factor in making the system acceptable to local managements.

\subsection{Accounting}

This list is appended to the site's {\tt gridmap} file with the 
generic userid prefix {\tt babar}.
Appending the list has the desirable effect that if a user has a specific
account at the site, and this is put into the {\tt gridmap} file by 
other means, then a grid request for their DN will take the specific
account rather than the generic one.

The generic account system \cite{generic-ref} provides a pool of accounts
({\tt babar01, babar02...babar99} or whatever)
and the user is mapped onto the first free pool account.
This gives outsiders the facility to use the site, while providing
accountability: if a particular account behaves (inadvertantly 
or deliberately) in an anti-social way, the DN that was given this 
account is known.
Furthermore these accounts can, if desired, be given lower priorities and
privileges than local ones.

If a DN has finished with the account and then submits a subsequent request,
it will be given the same account as last time. This is 
useful for retrieving
output from jobs.

We thus have a system of authorisation and accounting which scales,  
i.e. for $M$ users at $N$ sites it requires $N + M$ transactions
rather than $N \times M$, and yet which enables local users to have priority.
The system has proved easy to operate and is reliable.

\section {DATA LOCATION}

Data in BaBar is organised into runs, sets of approximately 600,000 events 
recorded by the experiment. When a run is finished the run number is 
incremented and the next run started.
The mapping of datafiles to runs is many-to-one:
each file contains the events from a particular run, there are many such
files containing different processing and selection stages but for each
(complete) processing specification there is a unique file for each run.
All the necessary metadata information can thus be contained in the
dataset name, which includes the run number, processing version, selection
program version, selection type and other information\cite{skimdata-ref}.

The data catalog is maintained centrally at SLAC, and is thus also the
metadata catalog: each filename contains a full and
systematic description of what it contains.  

Each site contains a copy of the main metadata catalog, updated nightly.
It also contains an extra flag for each file which is set if a local
copy exists.  This is maintained by the {\tt SkimTools} 
facilities \cite{skimtools-ref}
which are used, for example, to request datasets fulfilling certain
criteria to be copied from SLAC to the local site if they do not exist.

The name of each file is the same in all sites. To accommodate the fact
that different sites have different disk organisations, each file name
starts with the {\tt unix} symbolic variable {\tt \$BFROOT}. This is 
defined in a login script to point to an appropriate node in the file system
for that particular site.

This system does not map particularly usefully onto the `replica catalog'
concept \cite{replicacatalog-ref}. For this reason, and because of the problems of
speed and usability then being encountered with the general {\tt ldap} based
replica catalog of EDG, we decided to develop our own system for these
specific purposes.

This catalog can be queried with the {\tt skimData} command, which 
responds to requests for lists of the files satisfying  user-defined criteria
that exist locally. This list of files is actually produced as a 
series of {\tt tcl} commands, and hence known as `the {\tt .tcl} files',
and for convenience in splitting a complete analysis into jobs the
{\tt skimData} command is usually used to produce a number of {\tt .tcl}
files.  A typical analysis might involve $\approx 100$ such {\tt .tcl} files,
each containing a list of $\approx 100$ data files.

\subsection{Data and the Demonstrator}

For the demonstrator, use was made of the fact that {\tt skimData} can
respond to requests from a remote site.  The user specifies the file selection
criteria through as simple Web interface Perl/cgi script. 
They then specify, through a similar script, a list of sites in priority order.
{\tt skimData} is then invoked to produce {\tt .tcl} files for all runs
available at the first site.  It is then invoked for the second site, using the
list of runs found at the first site as a set of runs 
for which information is not desired,
(a facility originally provided to avoid undesirable data and hence
called `badruns')
 generating a second set of {\tt .tcl} files.
This process is then repeated for all sites specified.

\subsection{ Data and {\tt gsub}}

Although this method worked, the time taken for remote {\tt skimData} queries
could be inconveniently long (typically a few minutes).  The method was
therefore improved by adding to the catalog a list of availability at all sites.
This is maintained gathering the 
information by a nightly {\tt cron} job. For convenience it could be
maintained centrally and copied by local sites as it is done for the
rest 
of the
metadata. The system this scales for $N$ sites as
$2N$ 
rather 
than $N^2$.
The information is not quite up to date but this is not a practical
drawback.

This produces a single {\tt .tcl} file, and an index file which
describes which runs are available at which sites. Another Perl/cgi
script is used to divide these into {\tt .tcl} files to be run at each site
according to rather simple user criteria.

\section {THE INPUT SANDBOX}

A BaBar analysis job is performed by running a binary which takes as
its first and only argument the name of a {\tt .tcl} file. 
Typically it takes the form {\tt BetaApp myAnalysis.tcl},
where {\tt myAnalysis.tcl} is a small file which contains specific
user instructions; it sources one of the {\tt .tcl} files produced
by {\tt skimData} to locate the event data it is to run on, and also 
various standard BaBar {\tt .tcl} files, which in turn source other similar
{\tt .tcl} files.

In the standard (pre-Grid) system, such a job is run from a `working directory'
within the standard BaBar environment, i.e. various symbolic links and
environment variables are provided to all the files that will be needed
\cite {workdir-ref}. This
includes the standard BaBar {\tt .tcl} files and also some files used in
processing (such as tabulated efficiencies of Particle Identification.)

These other (non-{\tt tcl}) files pose a particular problem as there 
is no easy way of knowing which files will be required.  An analysis
may run satisfactorily in a test job, but in production there is nothing
to stop a low
probability branch in the analysis requiring the reading of an
unforeseen file.  There is a collaboration policy that the
information contained in such files should be moved to a central file
(the conditions database), but this is not yet complete.

It would be possible to build a system which relied on such an environment
also being available at the remote site. However we felt this to be unduly
restrictive - it requires not just that {\em a} BaBar environment be available,
but that {\em the} corresponding BaBar environment, which changes from
release to release, be available. So these files ({\tt .tcl} and others) must
be provided in the input sandbox for the job.

To run a job one needs
\begin{itemize}
\item The Event Data
\item The binary
\item The {\tt skimData} {\tt .tcl} file
\item The {\tt myAnalysis.tcl} file
\item The other {\tt .tcl} files
\item The other non-{\tt .tcl} files
\end{itemize}
Only the first does not present a problem. 
We adopted throughout the philosophy of taking the job to the data, rather
than moving the data to the job.
From the way the {\tt .tcl} files have been produced, the data 
will be present and accessible with the appropriate definition 
of {\tt \$BFROOT}.

\subsection{Inputs and the demonstrator}

As the demonstrator used a unique binary, this was compiled
(at SLAC) copied to a location (at Manchester) where it was accessible
via http. This was then copied (sucked) to the remote site at the
start of each job. 

The {\tt skimData.tcl} file for the job was sent in the input stream
as part of the input generated by the Perl/cgi script.

The {\tt myAnalysis.tcl} file was then expanded, i.e. each sourced
{\tt .tcl} file was replaced by the actual contents, iteratively until
the file was self-complete.  This was a major task to do by hand,
though a facility has now been provided to do this automatically, if necessary.
This large {\tt .tcl} file was then shipped to the remote sites in a similar manner to the binary,
as were the few but large non-{\tt tcl} files the standard job requires.

\subsection{Inputs and the {\tt gsub} command}

In the second version we use AFS to circumvent the input sandbox problem.
The user's working directory must be within an AFS filesystem
at a site running a {\tt gsiklogd} server. An AFS filesystem is not a problem as 
many BaBar working environments use, or have the option of using, AFS based
user directories (e.g. SLAC and RAL). The second is a requirement which
does require the co-operation of the system manager, but there are no
major problems associated with it.

At the start of each 
job {\tt gsiklog} is run (as a client) at the remote site. The binary
is copied if it is not available on the system, so this does not
restrict the choice of target sites.  This essentially provides an 
AFS token on the authority of the Grid proxy, which is used to run the job.
The job can then change directory to the user's working directory, and
all the links are available. 
All the {\tt .tcl} files, and the non-{\tt tcl} files, and the binary
are then available.
Some initialisation of environment variables is required but can readily be done
through standard login scripts - the only point where care is needed is 
to avoid confusion between the {\tt \$BFROOT} appropriate for the 
remote site and for the user's home site.

AFS access is well known to be slow and inappropriate for large
data transfers, owing to the time taken due to local caching etc. This
is not a major drawback, as these files are small and read only once 
during for each node. The data itself is read locally, using NFS.

\section{JOB SUBMISSION}

\subsection{Job submission for the Demonstrator}

Having set up the {\tt .tcl} files, or before, the user creates a
grid proxy on their local platform. This is then uploaded into the 
web server. This temporarily delegates the user's identity to the web
server. Using an unmodified web browser with the user's certificate,
the user can then instruct the server to perform remote job submission 
and other operations on the grid.

The server dispatches a set of jobs, one for each of the  {\tt .tcl} files.
For convenience in data collection, the set of jobs for each site is called
superjob, and each superjob starts with a {\tt job0} which
copies the binary, the expanded Babar {\tt .tcl} file, and the identified
other files to the remote site making them available for the subsequent
jobs.  The set of all superjobs (the hyperjob) is the complete `job'
in the sense that it is often used.

  Some initialisation of environment
variables is required but can readily be done through standard login
scripts - the only point where care is need is in possible confusion
between the appropriate {\tt \$BFROOT} for the remote site and that for
the user's home site.

Individual jobs run at the remote sites, usually through submission to
{\tt PBS}. Their progress can be monitored and their log files retrieved,
though this is rather slow and incovenient other than for test purposes.

\subsection{Submitting jobs with the {\tt gsub} command}

The basic {\tt BetaApp myAnalysis.tcl} commands are sent by 
{\tt globus-job-submit}
to the appropriate remote sites (again, usually to the {\tt pbs} batch system)
in a simple wrapper that does the {\tt gsiklog} and {\tt cd} to the working directory.
Each is preceded by a {\tt globus-job-run -stage} to 
provide {\tt gsiklog} if necessary, and other simple tasks.

For each {\tt .tcl} file the user (or a simple script) gives 
the command {\tt gsub <site> BetaApp myAnalysis.tcl}.
There is no need to categorise them into hyper and superjob
collections.

\section{OUTPUT RETRIEVAL}

The log files are not of much interest. Each analysis job produces as useful
output a set of ntuple/histogram files in hbook (or ROOT) format. The job
{\tt jobnn} typically processes a {\tt .tcl} file {\tt data-nn.tcl} and
produces a file {\tt output-nn.root}.

\subsection{Output from the  Demonstrator}
As each job finishes it moves its output file (stored on the farm node)
to a directory {\tt <superjobid>} on the gatekeeper for that site,
and tars together all the jobs there to form a single file. Thus as jobs
start to finish, there is always one file on the site that contains all the
presently available output data.

The user - again from a web browser running Perl/cgi - can then invoke a
job which runs on the server and copies all the tarred superjob files 
back to a directory {\tt <hyperjobid>} using {\tt grid-ftp}. They are untarred
and tarred into a single directory, and renamed as necessary so that
missing/unfinished/failed jobs do not give holes in the sequence.
A link is provided to the total tarred file, and a specific MIME type assigned.
The browser has this MIME type specified such that when the link is clicked on,
the file is downloaded to the browser and 
untarred, and a small ROOT
program run to produce a set of histograms from the total results.

\begin{figure}
\includegraphics[width=65mm]{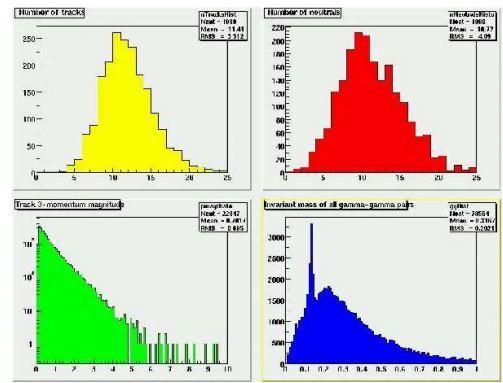}
\caption{Results from The Demonstrator}
\label{fr31}
\end{figure}

\subsection{Output from {\tt gsub} }

By contrast, this is absolutely trivial.
AFS also provides the output sandbox. The ntuple files are written
directly back to the users work directory and can be analysed there.

\section{CONCLUSIONS}

\subsection{Achievements}

We have found what appears to be a simple and stable solution to the
three A's problem. The Demonstrator has shown that simple grid tools can
be used to locate data across remote sites, run appropriate analysis
jobs, retrieve and combine the outputs.

The improved metadata catalog system and the {\tt gsub} command
 are providing 
system which has the flexibility for real users doing real analysis
in a real experiment

\subsection{Limitations}

There is no resource matching (except for the data). The user still has to make choices
about what sites to use, and balance loading appropriately.

The system is slow - each {\tt gsub} takes many seconds. For a typical analysis
 involving
several hundred jobs a script looping over {\tt gsub} can take a barely
acceptable time to process.  

A small fraction of jobs are lost. There is no bookkeeping system to take 
care of this, and providing one is not a simple add in.

\subsection{Outlook}

The {\tt gsub} command will be made available within BaBar. Whether it
is picked up and exploited by users will be an interesting test.
The {\tt globus-job-submit}  commands will be replaced by {\tt edg-job-submit}
as soon as the system becomes stable and supported.

\begin{acknowledgments}
The authors wish to thank 
Dr Jenny Williams who expanded the {\tt .tcl} file for the demonstrator.
\end{acknowledgments}


\end{document}